\documentclass[preprint,aps,showpacs,nofootinbib]{revtex4}

\usepackage{graphicx}
\usepackage{epsfig,amssymb,amsmath,color,epstopdf}

\def\e{\epsilon}
\def\f{\phi}

\newcommand{\beq}{\begin{equation}}
\newcommand{\eeq}{\end{equation}}
\newcommand{\bea}{\begin{eqnarray}}
\newcommand{\eea}{\end{eqnarray}}
\newcommand{\bear}{\begin{array}}
\newcommand {\eear}{\end{array}}
\newcommand{\bef}{\begin{figure}}
\newcommand {\eef}{\end{figure}}
\newcommand{\bec}{\begin{center}}
\newcommand {\eec}{\end{center}}

\def\REF#1{(\ref{#1})}
\def\GEV#1{10^{#1}{\rm\,GeV}}

\def\lrfp#1#2#3{ \left(\frac{#1}{#2} \right)^{#3}}

\begin{document}
\draft
\tighten
\preprint{
TU-960,~
IPMU14-0063
}
\title{\large \bf
Running Spectral Index from
Large-field Inflation
\\ with Modulations Revisited 
}
\author{
    Michael Czerny\,$^{a\,\star}$\footnote[0]{$^\star$ email: mczerny@tuhep.phys.tohoku.ac.jp},
    Takeshi Kobayashi,$^{b,\,c\,\ast}$\footnote[0]{$^\ast$ email: takeshi@cita.utoronto.ca},
    Fuminobu Takahashi\,$^{a,\,d\,\dagger} $\footnote[0]{$^\dagger$ email: fumi@tuhep.phys.tohoku.ac.jp}
    }
\affiliation{
    $^a$ Department of Physics, Tohoku University, Sendai 980-8578, Japan \\
    $^b$Canadian Institute for Theoretical Astrophysics, University of Toronto, \\ 
             60 St. George Street, Toronto, Ontario M5S 3H8, Canada \\
    $^c$Perimeter Institute for Theoretical Physics, \\ 
            31 Caroline Street North, Waterloo, Ontario N2L 2Y5, Canada\\
    $^d$ Kavli IPMU, TODIAS, University of Tokyo, Kashiwa 277-8583, Japan
    }
%\date{\today}

\vspace{2cm}

\begin{abstract}
We revisit large field inflation models with modulations in light of the recent discovery of the
primordial B-mode polarization by the BICEP2 experiment,  which, when combined with 
the {\it Planck} + WP + highL data, gives a strong hint for additional suppression 
of the CMB temperature fluctuations at small scales.
Such a suppression can be explained by a running spectral index.
In fact, it was pointed out by two of the present authors (TK and FT) that the existence of
both tensor mode perturbations and a sizable running of the spectral index is
a natural outcome of large inflation models with modulations such as axion monodromy inflation. 
We find that this holds also in the recently proposed multi-natural inflation, in which the inflaton potential consists
of multiple sinusoidal functions and therefore the modulations are a built-in feature.
\end{abstract}
\maketitle

\section{Introduction}
The BICEP2 experiment detected the primordial B-mode polarization of the cosmic microwave 
background (CMB) with very high significance~\cite{BICEP2}, giving a very strong case
for inflation~\cite{Guth:1980zm,Linde:1981mu}.
The inflation scale is determined to be
\bea
\label{B-Hr}
H_{\rm inf} &\simeq& 1.0 \times \GEV{14} \lrfp{r}{0.16}{\frac{1}{2}},\\
r &=& 0.20^{+0.07}_{-0.05} ~~(68\%{\rm CL}),
\eea
where $H_{\rm inf}$ is the Hubble parameter during inflation, and $r$ denotes the tensor-to-scalar ratio. 
The preferred range of $r$ is modified to $r = 0.16^{+0.06}_{-0.05}$,
after subtracting the best available estimate for foreground dust. The BICEP2 result strongly suggests
large-field inflation occurred, and  by far the simplest model
is the quadratic chaotic inflation~\cite{Linde:1983gd}.\footnote{ 
For various large-field inflation models and their concrete realization in the standard model as well as supergravity and superstring theory,
see e.g.~\cite{Freese:1990ni,Murayama:1992ua,Kawasaki:2000yn,Kallosh:2007ig,Silverstein:2008sg,McAllister:2008hb,Kaloper:2008fb,Takahashi:2010ky,
Nakayama:2010kt,Nakayama:2010sk,Harigaya:2012pg,Croon:2013ana,Nakayama:2013jka,Nakayama:2013nya,Czerny:2014wza,Czerny:2014xja,Nakayama:2014-HCI,Murayama:2014saa}}
The discovery of the tensor mode 
perturbations is of significant importance not only for cosmology but also for particle physics,
because  the suggested inflation energy scale is close to the GUT scale.
If the primordial B-mode polarization is measured with better accuracy by the Planck satellite and
other ground-based experiments, it will pin down the underlying inflation model, providing invaluable 
information on the UV physics such as string theory. 

The BICEP2 data, when combined with the {\it Planck}+WP+highL data, gives a strong hint for some additional
suppression of the CMB temperature fluctuations at small scales~\cite{BICEP2}. 
 This is because the large tensor mode perturbations
also contribute to the CMB temperature fluctuations at large scales, which causes the tension on the relative
size of scalar density perturbations at large and small scales. 
The suppression of the density perturbations at small
scales can be realized by e.g. (negative)
running of the spectral index, hot dark matter, etc.\footnote{
Before the BICEP2 results,
there was a hint for the presence of hot dark matter, such as sterile neutrinos~\cite{Wyman:2013lza,Hamann:2013iba, Battye:2013xqa}.
Non-thermally produced axions are also an interesting candidate~\cite{Jeong:2013oza,Higaki:2014zua}.
}  In this letter we focus on the running spectral index as a solution 
to this tension.

The spectral index of the curvature power spectrum ${\cal P_R}$
is defined by 
\bea
n_s(k)-1 &=& \frac{d \ln {\cal P_R}(k)}{d \ln k}, 
\eea
and the running of the spectral index is obtained as the differentiation
of $n_s$ with respect to $\ln k$. 
The preferred range of the running spectral index and its statistical significance are not given in \cite{BICEP2}.
Since the combination of the {\it Planck}+WP+highL data constrains the running as~\cite{Ade:2013zuv}
%%
%\bea
$d n_s/d \ln k = - 0.022 \pm 0.010~(68\% {\rm CL})$,
%\eea
%%
we expect that, once the BICEP2 data is combined, non-zero values of  $d n_s/d\ln k \approx -0.02 \sim -0.03$ will be 
suggested with strong significance.  
As a reference value, we will assume that the running is
approximately given by $d n_s/d\ln k  \sim -0.025$ over the observed cosmological scales,
but the precise value is not relevant for our purpose.\footnote{Note
that both the spectral index and its running are usually 
evaluated at a pivot scale, and the running is assumed 
to be scale-independent  in the MCMC analysis of the CMB data~\cite{Ade:2013zuv}.  
On the other hand, there is no firm ground to assume that they are completely scale-independent, and in fact,  they do depend
on scales in many scenarios. Therefore, the comparison between theory and observation must be done carefully,
and a dedicated analysis to each theoretical model would be necessary to
deduce some definite conclusions.
We also note that the joint analyses of the $Planck$ and BICEP2 datasets
have been performed in recent works such as~\cite{Abazajian:2014tqa},
see Eq.~(\ref{eq22}).}

In a single-field slow-roll inflation model with a featureless
potential, the running of the spectral index is of second 
order in the slow-roll parameters, and therefore of order
$10^{-3}$. Thus, it is a challenge to explain a running 
as large as $d n_s/d\ln k \sim - 0.025$.
For various proposals on this topic, see e.g.~Refs.~\cite{Chung:2003iu,Cline:2006db,Espinosa:2006pb,Joy:2007na,Joy:2008qd,Kawasaki:2003zv,Yamaguchi:2003fp,Easther:2006tv}.
In particular, \cite{Easther:2006tv} pointed out that a large negative
running that is more or less constant over the observed cosmological
scales would quickly terminate inflation within 
$N \lesssim 30$ in terms of the e-folding number.
However, it should be noted that such a discussion is based on the
assumption that the inflaton potential is  expanded in the Taylor series of the inflaton field with finite truncation. 
In fact,  it is possible to realize the running spectral index in 
simple single-field inflation models. In Ref.~\cite{Kobayashi:2010pz},
two of the present authors (TK and FT)  showed that a sizable running spectral
index can be realized without significant impact on the overall behavior
of the inflaton if there are small modulations on the inflaton potential.
(See also~\cite{Feng:2003mk} for related work.)
Here, in order for the inflaton dynamics to be locally affected by the
modulations, the inflaton field excursion must be relatively large as in the
large-field inflation. Therefore, both the tensor mode and the running
spectral index are a natural outcome of the large field inflation with
modulations.
Examples such as monomial inflaton potentials ($V = \lambda \phi^n$) with superimposed
periodic oscillations were studied in~\cite{Kobayashi:2010pz}.

In this letter we revisit the large-field inflation with modulations in light of the recent discovery of the primordial
B-mode polarization by the BICEP2. Along the lines of Ref.~\cite{Kobayashi:2010pz}, we study the recently 
proposed multi-natural inflation~\cite{Czerny:2014wza,Czerny:2014xja,Czerny:2014qqa} as an example. Interestingly, the existence of the periodic oscillations is a
built-in feature of multi-natural inflation.  We show that the negative running spectral index can be
realized without significant impact on the overall inflation dynamics, similar to the case studied before.
We will also show that  the predicted values of $(n_s, r)$ for quadratic
chaotic inflation and natural inflation can also be realized in multi-natural inflation.

\section{Implications of BICEP2  for inflation}
Before proceeding to the analysis, let us here briefly discuss the implications of the BICEP2 results for inflation. 
First, the inflaton field excursion during the last $60$ e-foldings exceeds the Planck scale, $M_p \simeq  2.4 \times \GEV{18}$, 
in a large field inflation model suggested by the BICEP2 result \REF{B-Hr}. One plausible way for having  
a good control of the inflaton potential over super-Planckian field values is to introduce a shift symmetry, 
under which the inflaton $\phi$ transforms as 
\bea
\label{shift}
\phi &\to& \phi + \alpha,
\eea
where $\alpha$ is a real transformation parameter. The shift symmetry needs to be explicitly broken
in order to generate the inflaton potential.
That is to say, the global continuous shift symmetry can be explicitly broken down to
a discrete one. 
For instance, the symmetry breaking could manifest itself as sinusoidal functions
in the inflaton potential.
If a single sinusoidal function dominates the inflaton potential, it is the natural inflation~\cite{Freese:1990ni}. 
On the other hand, if there are many sources for the explicit breaking,  the inflaton potential
may consist of multiple sinusoidal functions with different height and
periodicity. 
Such a case was investigated in~\cite{Czerny:2014wza,Czerny:2014xja}, which we refer to
as multi-natural inflation.
In this sense, the existence of small periodic oscillations is a built-in feature of the multi-natural inflation.
As we shall see below, a sizeable negative running spectral index can be generated in multi-natural inflation.

\section{Running spectral index from inflation with modulations}
\subsection{Basic idea}

Let us first review our basic idea from~\cite{Kobayashi:2010pz}
on generating a sizable running spectral index.
The point is that substructures in the inflaton potential
can affect the tilt and/or the running spectral index in a non-negligible way,
while not changing the overall behavior of the inflaton dynamics.

We consider an inflaton potential $V(\phi)$ with modulations, which 
 can be decomposed as
\beq
V(\phi)  = V_0(\phi) + V_{mod}(\phi),
\eeq
where the second term represents the modulations. 
We assume that the modulations  are so small  that the
slow-roll approximations 
\bea
  3 H \dot{\phi} &\simeq& - V'(\phi), \\
  3 H^2 M_p^2&\simeq& V(\phi), 
 \label{slow-roll}
\eea
remain  valid. (Here an overdot denotes a time derivative.)

The running in the spectral index is expressed in terms of the slow-roll parameters as
\bea
\frac{dn_s}{dlnk} \simeq -24\e^2 +16\e\eta - 2\xi ,
\label{8running}
\eea
where
\bea
\e \equiv \frac{M_p^2}{2}\left(\frac{V'}{V}\right)^2, \quad \eta \equiv M_p^2 \frac{V''}{V}, \quad \xi \equiv M_p^4 \frac{V'V'''}{V^2},
\eea
and primes denote derivatives with respect to the inflaton field
$\phi$. Thus, in order to generate a sizable
running, the third derivative of the inflaton potential must be large. 
At the same time, we consider the modulations to minimally 
affect the overall behavior of the inflaton dynamics.
%The modulations should generate sizable contribution to the running of the spectral index, 
%while keeping the overall behavior of the inflaton intact. 
To this end we require the following conditions for most of the inflaton field values:
\bea
\label{con1}
|V_0(\phi)|&\gg& |V_{mod}(\phi)|,\\
\label{con2}
|V_0'(\phi)| &\gtrsim& |V_{mod}'(\phi)|,\\
\label{con3}
|V_0''(\phi)| &\lesssim& |V_{mod}''(\phi)|,\\
\label{con4}
|V_0'''(\phi)| &\ll & |V_{mod}'''(\phi)|.
\eea
%
%The overall behavior of the inflaton is not changed by the modulations, if
We assume that
 the effect of the modulations $V_{mod}$ (and its derivatives) on the inflaton dynamics 
 should be negligibly small when averaged over a sufficiently long time or 
large field space.  One example for such modulations is a sinusoidal function.
If both $V_0$ and $V_{mod}$ are given by sinusoidal functions as in multi-natural inflation,
 $V_{mod}$ should have a shorter period, i.e., smaller decay constant to satisfy the
 above conditions.

The curvature perturbation power spectrum can be computed as\footnote{It
should also be noted that the formulae obtained using
the slow-roll approximations can break down when the higher order
derivatives of the inflaton potential become too large. In particular,
expressions such as (\ref{8running}) contain errors of order the
approximate results operated by
\begin{equation}
 \frac{1}{H}\frac{d}{dt} \simeq -M_p^2 \frac{V'}{V}\frac{d}{d \phi}.
\label{eq14}
\end{equation}
This quantity is smaller than unity in the cases studied in this letter,
as we consider examples where the oscillation amplitude of the spectral
index is larger than that of the running.
Hence we can invoke the slow-roll approximations.\label{foot4}}
\bea
{\cal P_R}(k) \simeq  \left.\frac{H^4}{4 \pi^2 \dot{\phi}^2}\right|_{k=aH}.
 \label{pr}
\eea
 Thus, the curvature power spectrum
receives modulations due to $\dot{\phi}^2 \propto |V'_0 + V_{mod}'|^2$.
Note that the modulations of $H$ due to $V_{mod}$ is subdominant, as long as 
condition (\ref{con1}) is met. 
On the other hand, the spectral index as well as the running are strongly affected
by the modulations if (\ref{con3}) and (\ref{con4}) are met. 
%
% It is clear that we can enhance the third derivative locally by introducing modulations satisfying (\ref{con1})-(\ref{con4}).

\subsection{Multi-natural inflation}
Now let us consider multi-natural inflation, where the inflaton potential consists of
multiple sinusoidal functions. For simplicity let us focus on the case of two sinusoidal functions with different heights
and periodicities. 
The potential takes  the form
\begin{equation}
  V(\f) = C - \Lambda_1^4\cos(\f/f_1) - \Lambda_2^4\cos(\f/f_2 + \theta),
  \label{eq:eni}
\end{equation}
where the decay constants $f_1$ and $f_2$ take different values.
$C$ is a constant that shifts the minimum of the potential to zero, and $\theta$ is a relative phase. 
The last term shifts the potential minimum from the origin to $\phi = \phi_{\rm min}$, and also modifies the potential shape.
This model is reduced to the original natural inflation in the limit of either $\Lambda_2 \rightarrow 0$ or $f_2 \rightarrow \infty$.
If we further take the limit of $\Lambda_1 \rightarrow \infty$ and $f_1 \rightarrow \infty$ while $\Lambda_1^2/f_1$ is kept constant,
the model is reduced to the quadratic chaotic inflation.

To simplify the notation we set $f_1 = f$ and $\Lambda_1 = \Lambda$, and
relate the parameters by,
\begin{align}
f_2 &= A f,\\
\Lambda_2^4 &= B \Lambda^4,
\label{eq:mn}
\end{align}
where $A$ and $B$ are real and positive constants. Because we are interested in small modulations to the inflaton potential, we choose
the parameters so that the conditions from (\ref{con1}) to  (\ref{con4}) are met. That is to say, $A$ and $B$ satisfy
\bea
A^3 \ll A^2 \lesssim B \lesssim A \ll 1.
\label{condAB}
\eea
%%

%Although in general $\Lambda_2$, $f_2$ and $\theta$ are arbitrary parameters, 
%we only investigate cases for which the second sinusoidal term gives relatively small perturbations, and
%the resulting potential is free of local minima so as to avoid the inflaton becoming trapped in a false
%vacuum.\footnote{
%In general there are many other local minima and maxima at different values of $\phi$. The stability of the vacuum at $\phi = \phi_{\rm min}$ is assumed in the following %analysis. 
%}

\begin{figure}[t!]
  \begin{center}
    \includegraphics[scale=0.31]{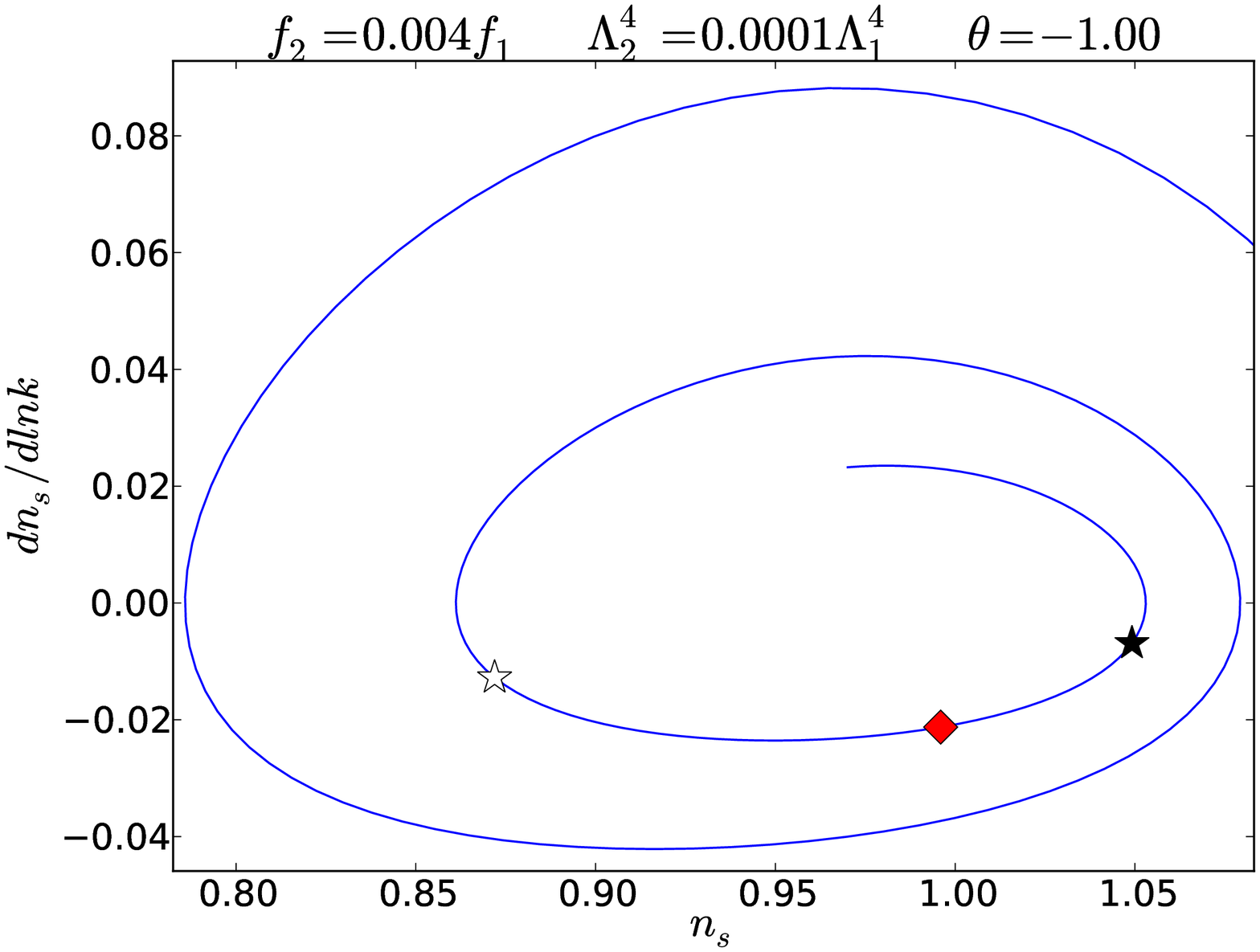}
    \includegraphics[scale=0.31]{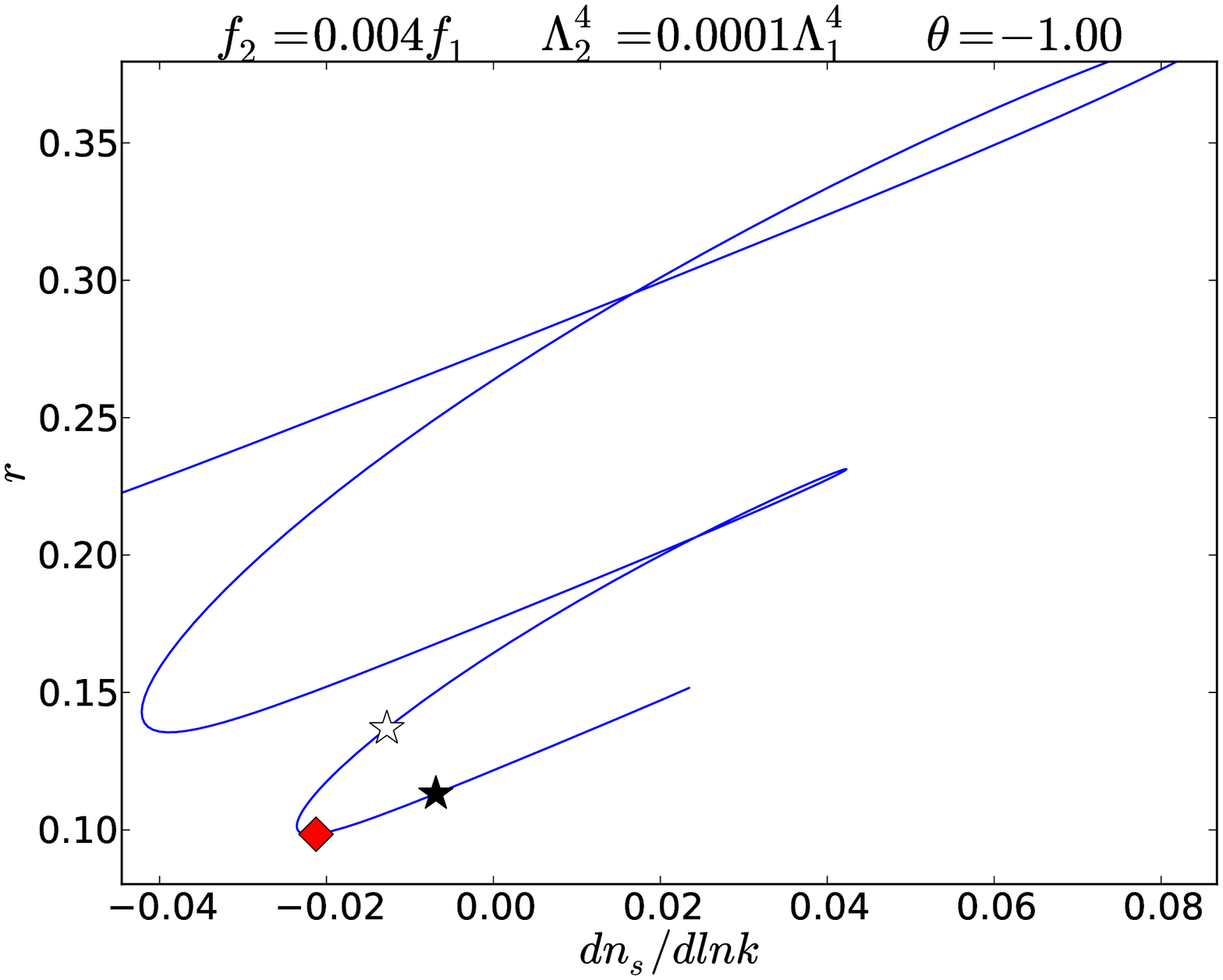}
    \includegraphics[scale=0.31]{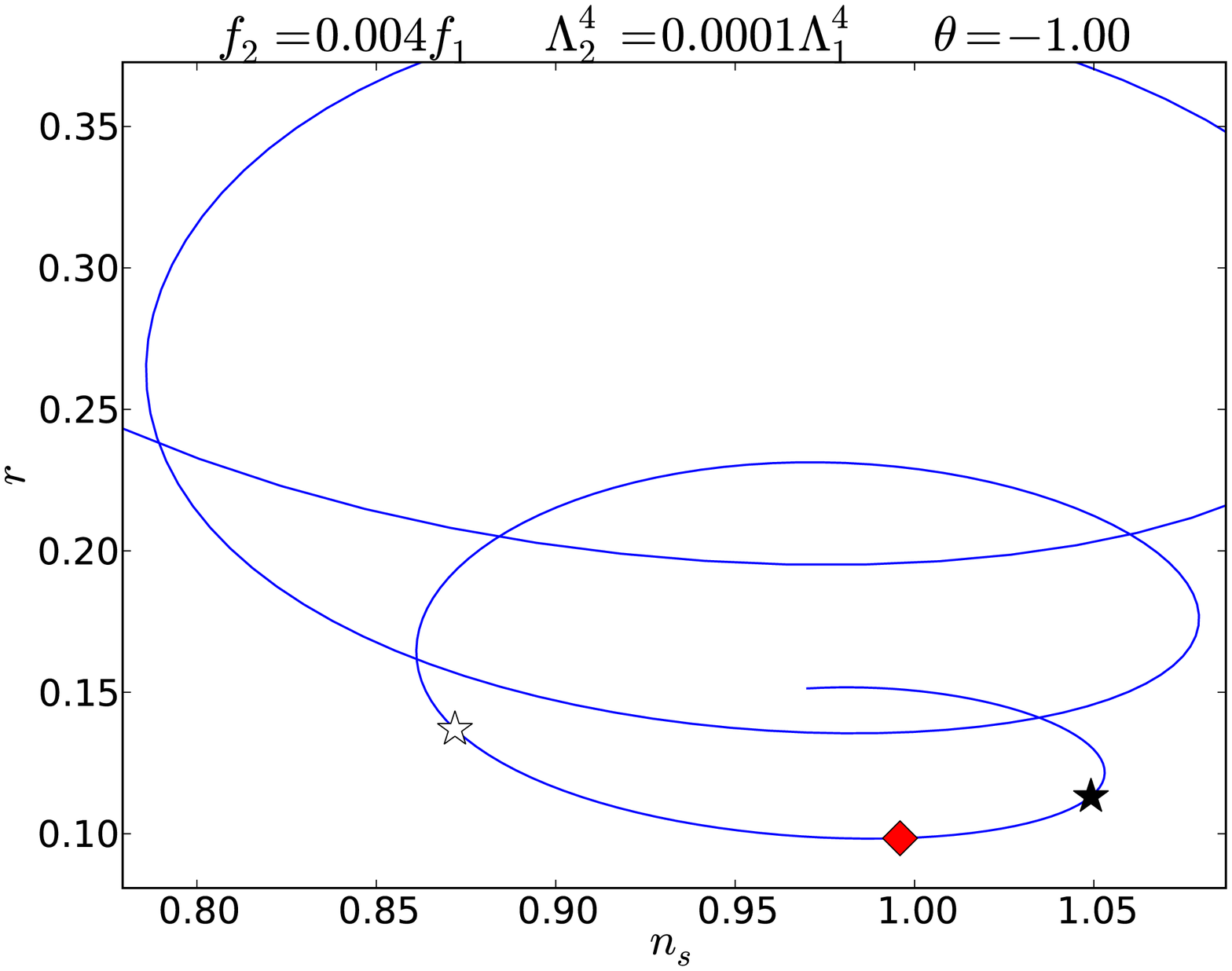}
  \end{center}
  \caption{Evolution of the spectral index $n_s$, its running  $d n_s/d \ln k$, and the tensor-to-scalar ratio for $f = 100M_p$,
 $A = 4 \times 10^{-3}$, $B = 1\times10^{-4}$, and $\theta = -1$.
  The red diamond denotes $N = 60$ e-folds before the end of inflation, whereas  the black (white) star corresponds to $N = 63 (55)$.}
  \label{fig:running1}
\end{figure}

\begin{figure}[t!]
  \begin{center}
    \includegraphics[scale=0.31]{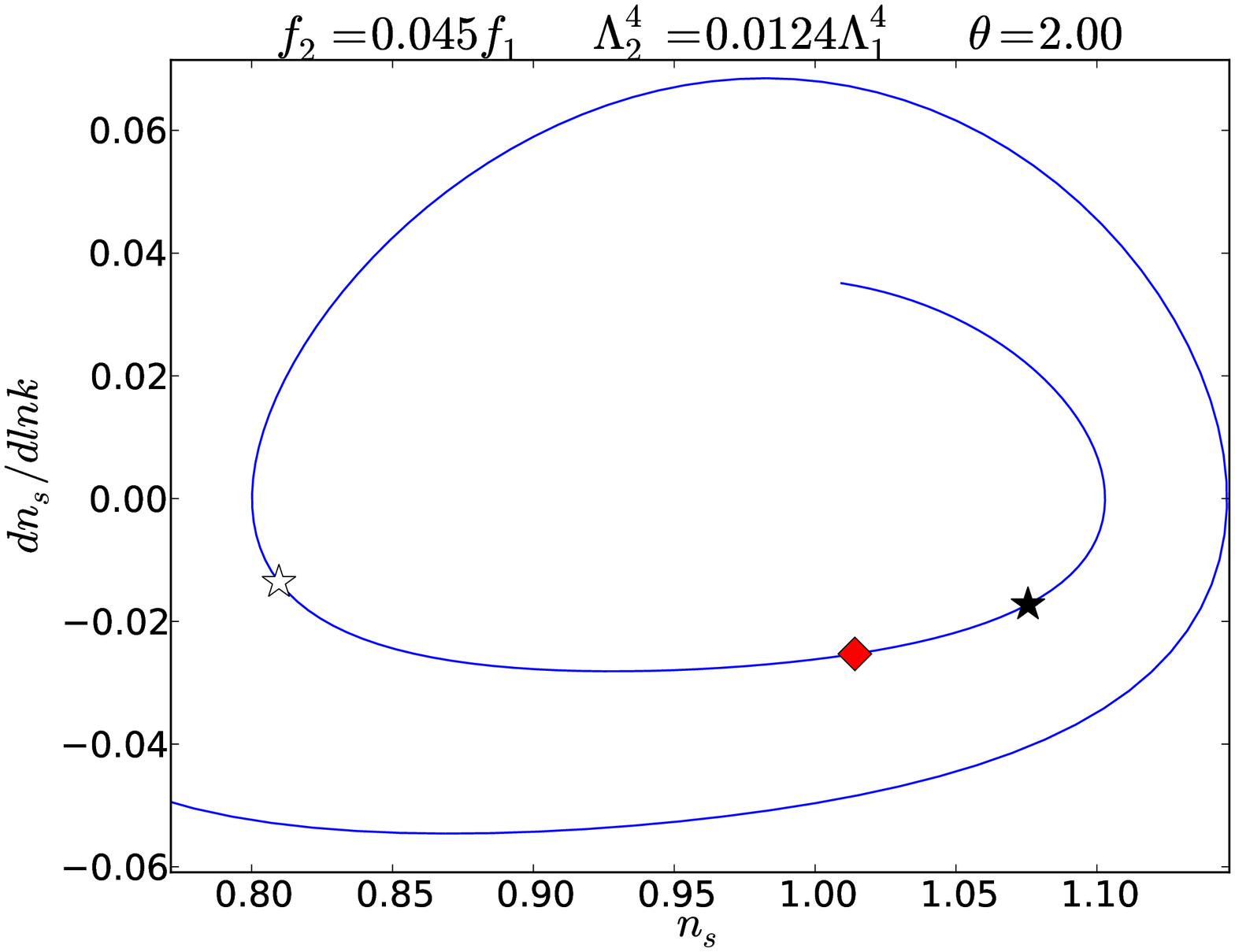}
    \includegraphics[scale=0.31]{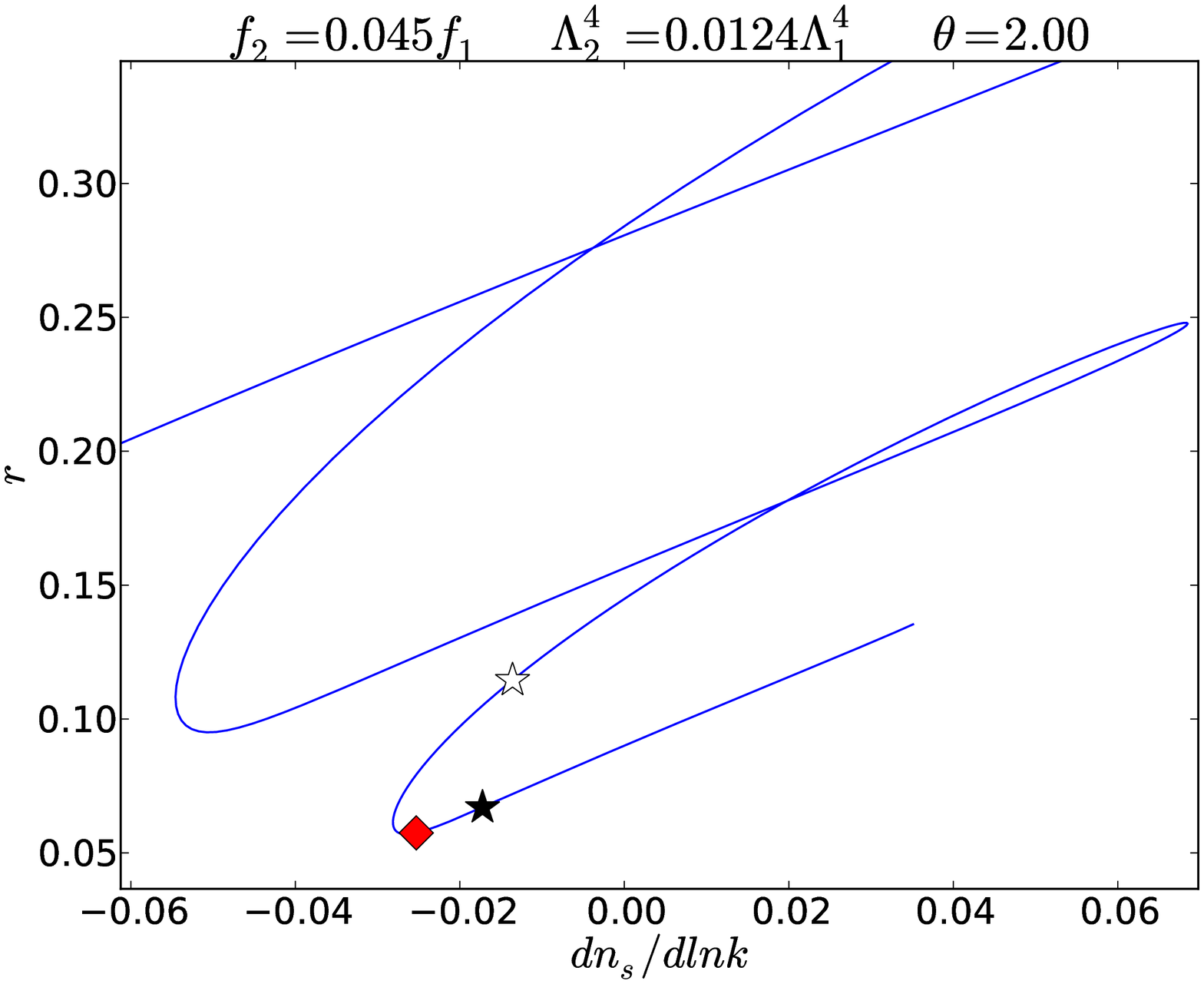}
    \includegraphics[scale=0.31]{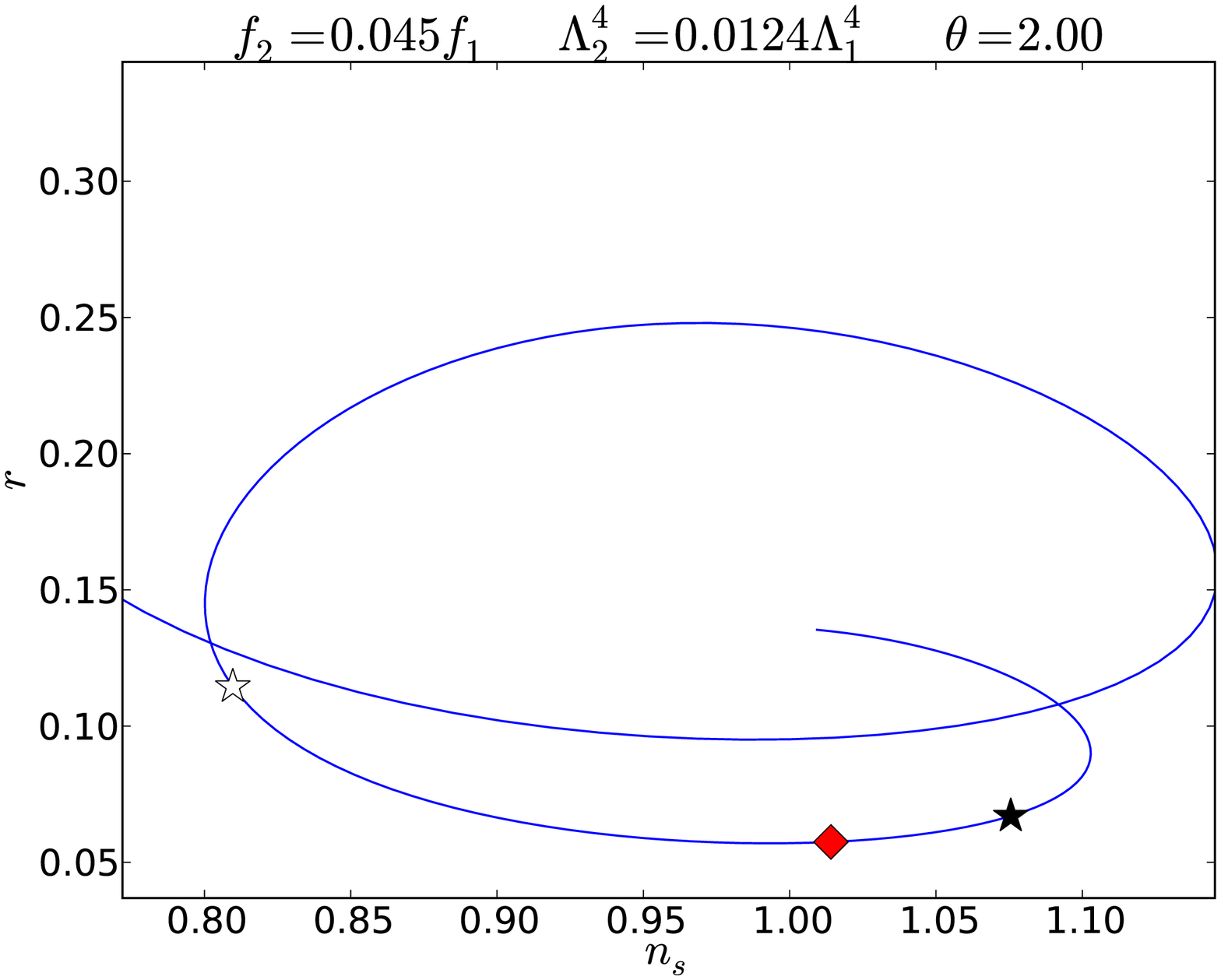}
  \end{center}
  \caption{Same as Fig.~\ref{fig:running1} but for $f = 10M_p$, $A=4.5\times 10^{-2}$, $B =  1.24\times 10^{-2}$ and $\theta = 2$. 
  The red diamond denotes $N=60$  e-folds before the end of inflation, whereas  the black (white) star corresponds to $N = 62 (54)$.}
  \label{fig:running2}
\end{figure}

Figs.~\ref{fig:running1} and~\ref{fig:running2} show the running as a function of $n_s$ and $r$ for $f = 100M_p$ 
and $f = 10M_p$, respectively.  In Fig.~\ref{fig:running1}  we take $A = 4 \times 10^{-3}$, $B = 1\times10^{-4}$, and $\theta = -1$, 
whereas in Fig.~\ref{fig:running2} we take $A=4.5\times 10^{-2}$, $B =  1.24\times 10^{-2}$ and $\theta = 2$. One can see that
the inequalities in (\ref{condAB}) are satisfied for these choices of the parameters. 
The case of $f=100 M_p$ is approximately the same as the case of the quadratic potential 
with small modulations~\cite{Kobayashi:2010pz}.
The potential height $\Lambda$ is fixed by the Planck normalization on the primordial density
perturbations~\cite{Ade:2013zuv}. Its typical value is approximately given by $\Lambda^2 \sim f\times  \GEV{13}.$
The predicted values of $n_s$, $r$, and the running $d n_s/d \ln k$ evolve clockwise around the curves and the tail of the curves correspond to $N = 70$ e-folds before the end of inflation.
In Fig.~\ref{fig:running1}, the black (white) star corresponds to $N = 63 (55)$. In Fig.~\ref{fig:running2}, the black (white) star corresponds to
$N = 62 (54$). In both cases, the red diamond is for $N = 60$.
Therefore a negative running spectral index can be realized in between the $N = 50$ to $60$ regime.
Thus, by taking the second sinusoidal term to be modulations in the potential, multi-natural inflation readily accommodates a negative running of $n_s$. 

In fact, there is a relation among the e-folding number during one period of modulations, $\Delta N$,  and
the oscillation amplitudes of $d n_s/d \ln k$ and $n_s$~\cite{Takahashi:2013tj}, given by
\bea
A_{dn_s/d\ln k} \sim \frac{2 \pi}{\Delta N} A_{n_s}.
\eea
This relation approximately holds in the above two examples. 

%Finally, in Fig.~\ref{fig:nsr_bicep2} we compare predicted values of ($n_s$, $r$) from multi-natural inflation with the most recent BICEP2 data. 
%For such a large value of $r$ reported by BICEP2, we see that multi-natural inflation can account for deviations in the predicted values of
%($n_s$, $r$) from natural inflation and the simple quadratic chaotic inflation model.

%\begin{figure}[t!]
%  \begin{center}
%    \includegraphics[scale=0.5]{multi_natural_bicep2.eps}
%  \end{center}
%  \caption{Predicted values of ($n_s$, $r$) for 3 values of $B$ (eq.~\REF{eq:mn}). Solid (dotted) lines correspond to N = 60 (N = 50) e-folds.}
%  \label{fig:nsr_bicep2}
%\end{figure}

\subsection{Comparison with Observational Constraints}
\label{subsecIIIC}

The combined $Planck$+WP+highL data constrains the running as
\begin{equation}
 dn_s / d \ln k = -0.022 \pm 0.010 \, \,  (68 \%),
\end{equation}
when allowing tensor fluctuations~\cite{Ade:2013zuv}. A recent
work~\cite{Abazajian:2014tqa} have performed a joint analysis of the  
$Planck$ and BICEP2 datasets, giving similar constraints,
\begin{equation}
 dn_s / d \ln k = -0.024 \pm 0.010 \, \,  (68 \%).
\label{eq22}
\end{equation}
The analyses assume a scale-independent running, while 
multi-natural inflation produces a running which itself oscillates with
respect to the wave number. However, when the oscillation period of the
modulations on the inflaton potential is large enough to incorporate
of order $10$ number of e-foldings, then the produced running is effectively
constant over the observed CMB scales. 
This is the case for the example parameters chosen in the previous
subsection; 
Figs.~\ref{fig:running1} and~\ref{fig:running2} show that 
over $\sim 10$ e-folds around the pivot scale, the running
varies by $\Delta (dn_s / d \ln k) \sim 0.01$, which is within the errors
of the constraints.
Hence the above running constraints should be valid for multi-natural
inflation as well. 
It would also be interesting to investigate cases with smaller oscillation
periods such that the running exhibits a strong scale-dependence within
the observed scales. (Though in such cases one should also consider the
applicability of the slow-roll approximations, see
Footnote~\ref{foot4}.)
More precise data from upcoming experiments may 
allow detailed investigations of the oscillation period of the modulations in
the inflaton potential.

\section{Discussions}

As we have pointed out in~\cite{Kobayashi:2010pz}, 
large field inflation with substructures in the inflaton potential 
entails large tensor perturbations as well as a running spectral index.
In this letter, we revisited large field models 
with modulations in the context of multi-natural inflation, where multiple
effects breaking the 
shift symmetry give rise to a superposition of sinusoidal functions to
the inflaton potential.
We focused on the interesting case where a hierarchy exists among the
periodicities of the sinusoidal oscillations, so that the model is a
large field model with superimposed periodic oscillations.
While the large field nature of the model produces large tensor mode
perturbations, the oscillations on the potential source a
running spectral index for the density perturbation spectrum.
We have seen that multi-natural inflation possesses rich phenomenology,
in particular, it produces a wide variety of values for $(r, \, n_s, \,
dn_s/d\ln k)$ depending on the relative size of the sinusoidal functions.

We also remark that 
large field inflation with modulations not only sources the running
spectral index for the density perturbations, but also for the tensor
perturbation spectrum.
The tensor spectral index and its running are given in terms of the slow-roll
parameters as
\begin{equation}
n_T = \frac{d\ln \mathcal{P}_T}{d \ln k }\simeq -2 \epsilon, 
\qquad
 \frac{dn_T}{d\ln k} \simeq -8 \epsilon^2 + 4 \epsilon \eta .
\end{equation}
Unlike~$n_s$, the tensor tilt depends only on~$\epsilon$ and thus 
the tensor running is set by $\epsilon$ and $\eta$. 
Therefore, one sees from the conditions (\ref{con1}) - (\ref{con4}) that
the running of the tensor tilt is smaller than that for the density
perturbations. Nonetheless, it is worth noting that a non-negligible 
$d n_s / d \ln k$ entails some amount of $d n_T / d \ln k$ as well.
This will become especially important when measuring the tensor tilt
by combining tensor observations at different scales,
such as when combining CMB experiments with direct observations of
gravity waves.
A simple extrapolation between widely different scales without considering the
possibility of the tensor running could lead to a misinterpretation
of the observational results; In particular, such a naive extrapolation
would give rise to an apparent violation of the slow-roll consistency relation~$r = -8
n_T$~\cite{Liddle:1993fq}, which holds locally at each scale in our case.

Upcoming experimental data are expected to verify whether there 
actually are sizable tensor mode perturbations and a running of the spectral
index. 
This will shed light on the substructure of the inflaton potential,
which should be directly tied to the underlying microphysics.

%%%%%%%%%%%%%%%%%%%%%%%%%%%%%%%%%%%%%
\section*{Acknowledgment}
%%%%%%%%%%%%%%%%%%%%%%%%%%%%%%%%%%%%%
This work was supported by Grant-in-Aid for  Scientific Research on Innovative
Areas (No.24111702, No. 21111006, and No.23104008) [FT], Scientific Research (A)
(No. 22244030 and No.21244033) [FT], and JSPS Grant-in-Aid for Young Scientists (B)
(No. 24740135 [FT]), and Inoue Foundation for Science [FT].
This work was also supported by World Premier International Center Initiative
(WPI Program), MEXT, Japan [FT].


\begin{thebibliography}{99}

 \bibitem{BICEP2} 
 BICEP2 Collaboration, http://bicepkeck.org
 

  %\cite{Guth:1980zm}
\bibitem{Guth:1980zm}
  A.~H.~Guth,
  %``The Inflationary Universe: A Possible Solution to the Horizon and Flatness Problems,''
  Phys.\ Rev.\  D {\bf 23}, 347-356 (1981);
  %\cite{Starobinsky:1980te}
%\bibitem{Starobinsky:1980te}
A.~A.~Starobinsky,
%``A New Type of Isotropic Cosmological Models without Singularity,''
Phys.\ Lett.\ B {\bf 91} (1980) 99;
%%CITATION = PHLTA,B91,99;%%
%\cite{Sato:1980yn}
%\bibitem{Sato:1980yn}
  K.~Sato,
  %``First Order Phase Transition of a Vacuum and Expansion of the Universe,''
  Mon.\ Not.\ Roy.\ Astron.\ Soc.\  {\bf 195}, 467-479 (1981).
  
  %\cite{Linde:1981mu}
\bibitem{Linde:1981mu}
A.~D.~Linde,
%``A New Inflationary Universe Scenario: a Possible Solution of the Horizon, Flatness, Homogeneity, Isotropy and Primordial Monopole Problems,''
Phys.\ Lett.\ B {\bf 108} (1982) 389;
%%CITATION = PHLTA,B108,389;%%
%\cite{Albrecht:1982wi}
%\bibitem{Albrecht:1982wi} 
  A.~Albrecht and P.~J.~Steinhardt,
  %``Cosmology for Grand Unified Theories with Radiatively Induced Symmetry Breaking,''
  Phys.\ Rev.\ Lett.\  {\bf 48}, 1220 (1982).
  %%CITATION = PRLTA,48,1220;%%

     %\cite{Linde:1983gd}
\bibitem{Linde:1983gd}
  A.~D.~Linde,
  %``Chaotic Inflation,''
  Phys.\ Lett.\ B {\bf 129}, 177 (1983).
  %%CITATION = PHLTA,B129,177;%%   
  
  
  
     %\cite{Freese:1990ni}
\bibitem{Freese:1990ni}
  K.~Freese, J.~A.~Frieman and A.~V.~Olinto,
  %``Natural inflation with Pseudo Nambu-Goldstone Bosons,"
  Phys.\ Rev.\ Lett.\ {\bf 65}, 3233 (1990).
  
  %\cite{Murayama:1992ua}
\bibitem{Murayama:1992ua} 
  H.~Murayama, H.~Suzuki, T.~Yanagida and J.~'i.~Yokoyama,
  %``Chaotic inflation and baryogenesis by right-handed sneutrinos,''
  Phys.\ Rev.\ Lett.\  {\bf 70}, 1912 (1993).
  %%CITATION = PRLTA,70,1912;%%
  %188 citations counted in INSPIRE as of 22 Mar 2014
  
  %\cite{Kawasaki:2000yn}
\bibitem{Kawasaki:2000yn} 
  M.~Kawasaki, M.~Yamaguchi and T.~Yanagida,
  %``Natural chaotic inflation in supergravity,''
  Phys.\ Rev.\ Lett.\  {\bf 85}, 3572 (2000)
  [hep-ph/0004243].
  %%CITATION = HEP-PH/0004243;%%
  %180 citations counted in INSPIRE as of 07 Mar 2014
  
%\cite{Kallosh:2007ig}
\bibitem{Kallosh:2007ig} 
  R.~Kallosh,
  %``On inflation in string theory,''
  Lect.\ Notes Phys.\  {\bf 738}, 119 (2008)
  [hep-th/0702059 [HEP-TH]].
  %%CITATION = HEP-TH/0702059;%%
  %172 citations counted in INSPIRE as of 22 Mar 2014
    
  %\cite{Silverstein:2008sg}
\bibitem{Silverstein:2008sg} 
  E.~Silverstein, A.~Westphal,
  %``Monodromy in the CMB: Gravity Waves and String Inflation,''
  Phys.\ Rev.\ D {\bf 78}, 106003 (2008)
  [arXiv:0803.3085 [hep-th]].
  %%CITATION = ARXIV:0803.3085;%%
  %146 citations counted in INSPIRE as of 29 Mar 2013
    
  %\cite{McAllister:2008hb}
\bibitem{McAllister:2008hb} 
  L.~McAllister, E.~Silverstein, A.~Westphal,
  %``Gravity Waves and Linear Inflation from Axion Monodromy,''
  Phys.\ Rev.\ D {\bf 82}, 046003 (2010)
  [arXiv:0808.0706 [hep-th]].
  %%CITATION = ARXIV:0808.0706;%%
  %128 citations counted in INSPIRE as of 29 Mar 2013
  
  %\cite{Kaloper:2008fb}
\bibitem{Kaloper:2008fb} 
  N.~Kaloper and L.~Sorbo,
  %``A Natural Framework for Chaotic Inflation,''  
Phys.\ Rev.\ Lett.\  {\bf 102}, 121301 (2009)  [arXiv:0811.1989 [hep-th]];
%%CITATION = ARXIV:0811.1989;%%  %35 citations counted in INSPIRE as of 25 Mar 2014
%
%\cite{Kaloper:2011jz}
%\bibitem{Kaloper:2011jz} 
  N.~Kaloper, A.~Lawrence and L.~Sorbo,
  %``An Ignoble Approach to Large Field Inflation,''  
JCAP {\bf 1103}, 023 (2011)  [arXiv:1101.0026 [hep-th]].  
%%CITATION = ARXIV:1101.0026;%%  %27 citations counted in INSPIRE as of 25 Mar 2014


  
%\cite{Takahashi:2010ky}
\bibitem{Takahashi:2010ky} 
  F.~Takahashi,
  %``Linear Inflation from Running Kinetic Term in Supergravity,''
  Phys.\ Lett.\ B {\bf 693}, 140 (2010)
  [arXiv:1006.2801 [hep-ph]].
  %%CITATION = ARXIV:1006.2801;%%
  %18 citations counted in INSPIRE as of 28 Mar 2013
  
  %\cite{Nakayama:2010kt}
\bibitem{Nakayama:2010kt} 
  K.~Nakayama, F.~Takahashi,
  %``Running Kinetic Inflation,''
  JCAP {\bf 1011}, 009 (2010)
  [arXiv:1008.2956 [hep-ph]];
  %%CITATION = ARXIV:1008.2956;%%
  %21 citations counted in INSPIRE as of 28 Mar 2013
  %17 citations counted in INSPIRE as of 28 Mar 2013
%\cite{Nakayama:2010ga}
%\bibitem{Nakayama:2010ga} 
%  K.~Nakayama, F.~Takahashi and ,
  %``General Analysis of Inflation in the Jordan frame Supergravity,''
  see also {\it ibid},
  JCAP {\bf 1011}, 039 (2010)
  [arXiv:1009.3399 [hep-ph]].
  %%CITATION = ARXIV:1009.3399;%%
  %8 citations counted in INSPIRE as of 28 Mar 2013

%\cite{Nakayama:2010sk}
\bibitem{Nakayama:2010sk} 
  K.~Nakayama, F.~Takahashi and ,
  %``Higgs Chaotic Inflation in Standard Model and NMSSM,''
  JCAP {\bf 1102}, 010 (2011)
  [arXiv:1008.4457 [hep-ph]];
  %%CITATION = ARXIV:1008.4457;%%


%\cite{Harigaya:2012pg}
\bibitem{Harigaya:2012pg} 
  K.~Harigaya, M.~Ibe, K.~Schmitz and T.~T.~Yanagida,
  %``Chaotic Inflation with a Fractional Power-Law Potential in Strongly Coupled Gauge Theories,''
  Phys.\ Lett.\ B {\bf 720}, 125 (2013)
  [arXiv:1211.6241 [hep-ph]].
  %%CITATION = ARXIV:1211.6241;%%
  %10 citations counted in INSPIRE as of 07 Mar 2014

  %\cite{Croon:2013ana}
\bibitem{Croon:2013ana} 
  D.~Croon, J.~Ellis and N.~E.~Mavromatos,
  %``Wess-Zumino Inflation in Light of Planck,''
  Physics Letters B {\bf 724}, , 165 (2013)
  [arXiv:1303.6253 [astro-ph.CO]].
  %%CITATION = ARXIV:1303.6253;%%
  %19 citations counted in INSPIRE as of 07 Mar 2014
  
  %\cite{Nakayama:2013jka}
\bibitem{Nakayama:2013jka} 
  K.~Nakayama, F.~Takahashi and T.~T.~Yanagida,
  %``Polynomial Chaotic Inflation in the Planck Era,''
  Phys.\ Lett.\ B {\bf 725}, 111 (2013)
  [arXiv:1303.7315 [hep-ph]];
  %%CITATION = ARXIV:1303.7315;%%
  %21 citations counted in INSPIRE as of 07 Mar 2014
%  %\cite{Nakayama:2013txa}
%\bibitem{Nakayama:2013txa} 
%  K.~Nakayama, F.~Takahashi and T.~T.~Yanagida,
  %``Polynomial Chaotic Inflation in Supergravity,''
  JCAP {\bf 1308}, 038 (2013)
  [arXiv:1305.5099 [hep-ph]].
  %%CITATION = ARXIV:1305.5099,;%%
  %17 citations counted in INSPIRE as of 07 Mar 2014
  
  %\cite{Nakayama:2013nya}
\bibitem{Nakayama:2013nya} 
  K.~Nakayama, F.~Takahashi and T.~T.~Yanagida,
  %``Chaotic Inflation with Right-handed Sneutrinos after Planck,''
  Phys.\ Lett.\ B {\bf 730}, 24 (2014)
  [arXiv:1311.4253 [hep-ph]].
  %%CITATION = ARXIV:1311.4253;%%
  %5 citations counted in INSPIRE as of 22 Mar 2014
  
  %\cite{Czerny:2014wza}
\bibitem{Czerny:2014wza} 
  M.~Czerny and F.~Takahashi,
  %``Multi-Natural Inflation,''
  arXiv:1401.5212 [hep-ph], to appear in Phys. Lett. B.
  %%CITATION = ARXIV:1401.5212;%%
  %1 citations counted in INSPIRE as of 07 Mar 2014

%\cite{Czerny:2014xja}
\bibitem{Czerny:2014xja} 
  M.~Czerny, T.~Higaki and F.~Takahashi,
  %``Axion Hilltop Inflation in Supergravity,''
  arXiv:1403.0410 [hep-ph], to appear in Phys. Lett. B.
  %%CITATION = ARXIV:1403.0410;%%

%\cite{Nakayama:2014-HCI}
\bibitem{Nakayama:2014-HCI} 
  K.~Nakayama and  F.~Takahashi,
arXiv:1403.4132 [hep-ph].
  %%CITATION = ARXIV:1403.4132;%%

%\cite{Murayama:2014saa}
\bibitem{Murayama:2014saa} 
  H.~Murayama, K.~Nakayama, F.~Takahashi and T.~T.~Yanagida,
  %``Sneutrino Chaotic Inflation and Landscape,''
  arXiv:1404.3857 [hep-ph].
  %%CITATION = ARXIV:1404.3857;%%
  %7 citations counted in INSPIRE as of 07 May 2014

%\cite{Wyman:2013lza}
\bibitem{Wyman:2013lza}
  M.~Wyman, D.~H.~Rudd, R.~A.~Vanderveld and W.~Hu,
  %``$\nu\Lambda$CDM: Neutrinos help reconcile Planck with the Local Universe,''
  Phys.\ Rev.\ Lett.\  {\bf 112}, 051302 (2014)
  [arXiv:1307.7715 [astro-ph.CO]].
  %%CITATION = ARXIV:1307.7715;%%
  %27 citations counted in INSPIRE as of 26 Feb 2014

%\cite{Hamann:2013iba}
\bibitem{Hamann:2013iba}
  J.~Hamann and J.~Hasenkamp,
  %``A new life for sterile neutrinos: resolving inconsistencies using hot dark matter,''
  JCAP {\bf 1310}, 044 (2013)
  [arXiv:1308.3255 [astro-ph.CO]].
  %%CITATION = ARXIV:1308.3255;%%
  %12 citations counted in INSPIRE as of 23 Dec 2013

%\cite{Battye:2013xqa}
\bibitem{Battye:2013xqa}
  R.~A.~Battye and A.~Moss,
  %``Evidence for massive neutrinos from CMB and lensing observations,''
  Phys.\ Rev.\ Lett.\  {\bf 112}, 051303 (2014)
  [arXiv:1308.5870 [astro-ph.CO]].
  %%CITATION = ARXIV:1308.5870;%%
  %15 citations counted in INSPIRE as of 26 Feb 2014
  
  %\cite{Jeong:2013oza}
\bibitem{Jeong:2013oza} 
  K.~S.~Jeong, M.~Kawasaki and F.~Takahashi,
  %``Axions as Hot and Cold Dark Matter,''
  JCAP {\bf 1402}, 046 (2014)
  [arXiv:1310.1774 [hep-ph], arXiv:1310.1774].
  %%CITATION = ARXIV:1310.1774;%%
  %2 citations counted in INSPIRE as of 12 Mar 2014
  
  %\cite{Higaki:2014zua}
\bibitem{Higaki:2014zua} 
  T.~Higaki, K.~S.~Jeong and F.~Takahashi,
  %``The 7 keV axion dark matter and the X-ray line signal,''
  arXiv:1402.6965 [hep-ph].
  %%CITATION = ARXIV:1402.6965;%%
  %8 citations counted in INSPIRE as of 12 Mar 2014
  
    %\cite{Ade:2013zuv}
\bibitem{Ade:2013zuv} 
  P.~A.~R.~Ade {\it et al.}  [Planck Collaboration],
  %``Planck 2013 results. XVI. Cosmological parameters,''
  arXiv:1303.5076 [astro-ph.CO].
  %%CITATION = ARXIV:1303.5076;%%
  %117 citations counted in INSPIRE as of 03 May 2013

%\cite{Abazajian:2014tqa}
\bibitem{Abazajian:2014tqa} 
  K.~N.~Abazajian, G.~Aslanyan, R.~Easther and L.~C.~Price,
  %``The Knotted Sky II: Does BICEP2 require a nontrivial primordial power spectrum?,''
  arXiv:1403.5922 [astro-ph.CO].
  %%CITATION = ARXIV:1403.5922;%%
  
%\cite{Chung:2003iu}
\bibitem{Chung:2003iu}
  D.~J.~H.~Chung, G.~Shiu and M.~Trodden,
  %``Running of the scalar spectral index from inflationary models,''
  Phys.\ Rev.\  D {\bf 68}, 063501 (2003)
  [arXiv:astro-ph/0305193].
  %%CITATION = PHRVA,D68,063501;%%

%\cite{Cline:2006db}
\bibitem{Cline:2006db}
  J.~M.~Cline and L.~Hoi,
  %``Inflationary potential reconstruction for a WMAP running power  spectrum,''
  JCAP {\bf 0606}, 007 (2006)
  [arXiv:astro-ph/0603403].
  %%CITATION = JCAPA,0606,007;%%

%\cite{Espinosa:2006pb}
\bibitem{Espinosa:2006pb}
  J.~R.~Espinosa,
  %``The running spectral index as a probe of physics at high energies,''
  arXiv:hep-ph/0605150.
  %%CITATION = HEP-PH/0605150;%%

%\cite{Joy:2007na}
\bibitem{Joy:2007na}
  M.~Joy, V.~Sahni and A.~A.~Starobinsky,
  %``A New Universal Local Feature in the Inflationary Perturbation Spectrum,''
  Phys.\ Rev.\  D {\bf 77}, 023514 (2008)
  [arXiv:0711.1585 [astro-ph]].
  %%CITATION = PHRVA,D77,023514;%%

%\cite{Joy:2008qd}
\bibitem{Joy:2008qd}
  M.~Joy, A.~Shafieloo, V.~Sahni and A.~A.~Starobinsky,
  %``Is a step in the primordial spectral index favored by CMB data ?,''
  JCAP {\bf 0906}, 028 (2009)
  [arXiv:0807.3334 [astro-ph]].
  %%CITATION = JCAPA,0906,028;%%
  
%\cite{Kawasaki:2003zv}
\bibitem{Kawasaki:2003zv}
  M.~Kawasaki, M.~Yamaguchi and J.~Yokoyama,
  %``Inflation with a running spectral index in supergravity,''
  Phys.\ Rev.\  D {\bf 68}, 023508 (2003)
  [arXiv:hep-ph/0304161].
  %%CITATION = PHRVA,D68,023508;%%
  
%\cite{Yamaguchi:2003fp}
\bibitem{Yamaguchi:2003fp}
  M.~Yamaguchi and J.~Yokoyama,
  %``Chaotic hybrid new inflation in supergravity with a running spectral
  %index,''
  Phys.\ Rev.\  D {\bf 68}, 123520 (2003)
  [arXiv:hep-ph/0307373].
  %%CITATION = PHRVA,D68,123520;%%
  

  
  %\cite{Easther:2006tv}
\bibitem{Easther:2006tv} 
  R.~Easther and H.~Peiris,
  %``Implications of a Running Spectral Index for Slow Roll Inflation,''
  JCAP {\bf 0609}, 010 (2006)
  [astro-ph/0604214].
  %%CITATION = ASTRO-PH/0604214;%%
  %72 citations counted in INSPIRE as of 12 Mar 2014
  

  %\cite{Kobayashi:2010pz}
\bibitem{Kobayashi:2010pz} 
  T.~Kobayashi and F.~Takahashi,
  %``Running Spectral Index from Inflation with Modulations,''
  JCAP {\bf 1101}, 026 (2011)
  [arXiv:1011.3988 [astro-ph.CO]].
  %%CITATION = ARXIV:1011.3988;%%

%\cite{Feng:2003mk}
\bibitem{Feng:2003mk} 
  B.~Feng, M.~-z.~Li, R.~-J.~Zhang and X.~-m.~Zhang,
  %``An inflation model with large variations in spectral index,''
  Phys.\ Rev.\ D {\bf 68}, 103511 (2003)
  [astro-ph/0302479].
  %%CITATION = ASTRO-PH/0302479;%%
  
  %\cite{Czerny:2014qqa}
\bibitem{Czerny:2014qqa} 
  M.~Czerny, T.~Higaki and F.~Takahashi,
  %``Multi-Natural Inflation in Supergravity and BICEP2,''
  arXiv:1403.5883 [hep-ph].
  %%CITATION = ARXIV:1403.5883;%%
  %18 citations counted in INSPIRE as of 07 May 2014
  
%\cite{Takahashi:2013tj}
\bibitem{Takahashi:2013tj} 
  F.~Takahashi,
  %``The Spectral Index and its Running in Axionic Curvaton,''
  JCAP {\bf 1306}, 013 (2013)
  [arXiv:1301.2834, arXiv:1301.2834 [astro-ph.CO]].
  %%CITATION = ARXIV:1301.2834,;%%
  %1 citations counted in INSPIRE as of 12 Mar 2014


%\cite{Liddle:1993fq}
\bibitem{Liddle:1993fq} 
  A.~R.~Liddle and D.~H.~Lyth,
  %``The Cold dark matter density perturbation,''
  Phys.\ Rept.\  {\bf 231}, 1 (1993)
  [astro-ph/9303019].
  %%CITATION = ASTRO-PH/9303019;%%


\end{thebibliography}
\end{document}